# Protein Adaptive Plasticity and Night Vision

## J. C. Phillips

Dept. of Physics and Astronomy, Rutgers University, Piscataway, N. J., 08854

## Abstract

Proteins appear to be the most dramatic natural example of self-organized network criticality (SONC), a concept that explains many otherwise apparently exponentially unlikely phenomena.   Adaptive plasticity is a term which has become much more specific as a result of recent physiological and genetic studies.   Here we show that the molecular properties of rhodopsin, the transmembrane protein associated with night vision, can be quantified species by species using the Moret-Zebende hydropathicity scale based on SONC.   The results show that long-range adaptive plasticity optimizes proximate species molecular functionality far more effectively than one would infer using only standard amino acid sequence (local similarity) tools such as BLAST for multiple alignments. These results should be universal, and they suggest new paths for analyzing and predicting protein functionality from amino acid sequences alone.

## Introduction

Amino acid (aa) sequences dominate protein data bases, and > 5000 aa sequences from hundreds of species are known for common proteins such as lysozyme and rhodopsin. The aa are chosen from a menu of twenty aa selected by nature, both for their ability to pack (or fold) well and for their ability to respond to stimuli and perform a wide diversity of functions.   Over the last 50 years crystallographers have determined many protein structures, establishing isostructural superfamilies whose functionality varies significantly between species and even within a species family. These homological variations are attributable to ~ 10% mutations of the aa sequences, but in spite of many mutational studies, the reasons for most of the mutations have often remained obscure.

As a first "top-down" step towards understanding these variations, one can divide the mutational effects into short-range and long-range. One can search for naturally occurring short-range mutations with standard Web-based multiple alignment search tools, such as BLAST, which can also be used to compare family and superfamily interspecies variations, and sometimes find unexpected similarities between segments of ~ 10 aa or less from functionally dissimilar proteins. However, typical protein sequences with ~ 300 aa often contain secondary structural subunits with ~ 20-30 aa, where short-range BLAST searches may be inconclusive.

Long-range interactions are primarily responsible for globular folding: hydrophilic aa tend to segregate to the surface, while hydrophobic aa locate inside the globule. Because so many protein crystal structures are known, one can average over many of them using Voronoi partitioning to construct polyhedra around each aa of a protein structure, and then determine its solvent accessible surface area (SASA, for a 2 A water molecule); at present this construction is the basis of the most popular hydropathicity scale (there are altogether dozens of hydropathicity scales, with most based on transference energies from water to organic solvents). The comparative advantage of the SASA scale is that it accurately summarizes known protein geometries, and does not involve unreasonable assumptions (such as those involved in scales based on aa transference energies: proteins evolved only in water!). However, because of the short-range character of the Voronoi partition, the SASA scale unavoidably mixes short- and long-range interactions.

Moret and Zebende (MZ) discovered an elegant, self-validating way to isolate long-range hydropathic interactions[1]. Selective evolution has optimized protein networks both for stability and functionality, and the SASA of such optimized, self-organized networks are expected to be near thermodynamic criticality. They found that for long protein segments of length 2N+1, with $4 \leq N \leq 17$, $\text{dlog(SASA(aa))/dlogN} = -\psi(aa)$, in other words, the decrease of SASA with increasing segment size asymptotically follows a fractal power law. That self-similar long wave length decrease reflects increasing overlap of Voronoi surface areas with increasing length, and it is obviously larger for interior hydrophobic aa than for surface hydrophilic aa. The success of this kind of global criticality suggests that it may be caused by the presence of a fractal Riemannian 20-dimensional curved metric in an evolutionarily averaged protein configuration space. The breakdown of the power law for $N < 4$ represents the effects of non-hydropathic short-range packing interactions.

Because the MZ exponents $\{\psi(aa)\}$ are dimensionless and are based on SONC, we expect them to be more accurate than other scales, but how much more accurate? Perhaps the best way to answer this question is through examples. Lysozyme $c$ (140 aa) is a good place to start, because it is known for many species from insects to humans, and has dual functionality (metabolic and antibiotic). Moreover, the chicken and human structures are identical to 0.5 A, yet there is 40% mutational sequence difference, which produces considerable "noise". Hydropathic profile analysis of lysozyme $c$ based on averaging $\psi(aa)$ over rectangular windows of width W = 3 produced superior results that easily explained dual functional trends for seven species[2].

Here we extend the analysis to membrane proteins, specifically rhodopsin (348 aa), the best studied transmembrane protein belonging to the Guanine Protein Coupled Receptor (GPCR) superfamily, the largest family of proteins in the human genome (800 members). GPCR proteins have characteristic heptad structures, with seven long (25aa), predominantly helical, transmembrane (TM) interior sections connected by exterior or surface extracellular (EC) and cytoplasmic (CP) loops[3]. Their amino acid sequences form the largest database for protein-membrane interactions, and they perform a variety of functions: rhodopsin (visual signaling), adrenopsin (stimulative), adenopsin (metabolic), etc.

Given their known heptad structures, and the locations of their chemical receptors (retinal, adrenalin, etc.) between transmembrane (TM) interior sections[3], it is clear that functionality will depend on larger values of segmental lengths W in GPCR proteins than in lysozyme. Studies of some of these GPCR show satisfactory hierarchical separation of properties for $9 \leq W \leq 47$, which dramatically confirms the ability of the MZ scale to separate short-and long-range contributions of functionality, and shows that these functional differences are primarily effectuated by water-protein interactions[4]. To convolute $\psi$(aa) on this large W scale, one might try Fourier transforms, but this encounters phase problems associated with TM medium-length helical internal segmentation and helix capping (the variability of helical/non-helical boundaries[5,6]), so we naturally turn instead to the phase-independent variance. If we simply use the variance relative to an average over the entire protein, we obtain good results, but we can do even better: we can average separately over transmembrane, extracellular and cytoplasmic regions as listed in Uniprot, and calculate the variance by averaging over all three regions with physically distinct background averages for each region. We refer to this variance as the separated roughness form factor $\mathscr{R}^*(W)$; the idea is that evolution and plastic adaptivity should be reflected in species trends of $\mathscr{R}^*(W)$, which nature refines in order to optimize the night signaling ability of rhodopsin.

The reader may or may not find this credible, but this separated construction was only our second one for GPCR proteins (after the plain variance, tested first because EXCEL has a convenient variance tool), and it has turned out to be fully successful (no adjustable parameters!), although using it requires an EXCEL macro[4]. Separated variances, and profiles of MZ window averages, have shown several striking interspecies effects that are unrecognizable by multiple sequence alignment. We now proceed to discuss several such effects.

**High Pressure Environments**

Evolution dominates inter-species comparisons based on short-range sequence similarity, and if one utilizes human rhodopsin as a benchmark, most of the resulting similarities follow satisfactory evolutionary patterns, as expected from phylogenetic analysis (~ 40,000 papers on evolution* AND phylogenetic* in the last 25 years)[7]. However, for sequences alone short-range similarity can be misleading. An obvious problem is that the similarity of A and B does not establish an evolutionary direction, and multiple alignment matrices can be even less clear. One might have expected that human $\mathscr{R}^*(W)$ < other species $\mathscr{R}^*(W)$, as evolution has fully optimized the human rod and cone vision system (only humans and a few other primates have trichromatic cone vision), and this was largely confirmed, as a smoother sequence should transmit signals more efficiently. However, beyond evolution there is a second effect which we discuss first, that of high pressure on deep sea marine species. At high pressures, smoother sequences may be functionally more effective and less subject to damage by pressure fluctuations.

An elegant example that tests this idea is river and deep sea lamprey rhodopsins, which are encoded by a single gene[8,9]. The aa of the two species differ at 29 out of 353 sites, and three of these have been identified as responsible for causing a blue shift in the rhodopsin absorption spectra for adaptation to the blue-green photic environment in deep water[10]. This leaves twenty six aa replacements to be explained. Structurally 20 out of 171 differences are located in TM regions, and 9 out of 182 in EC and CP loop sites. The predominance of TM substitutions is understandable, as the increased deep-water pressure constrains internal pore-confined TM motion more than surface loop motion. However, when we compare BLAST similarity and $\mathscr{R}^*(W)$ for the two rhodopsin adaptations with human rhodopsin, quite a different picture emerges. For $W \leq 25$ (TM length L or shorter), the differences are small, but at $W = 47$ (2L), they are large (Table I).

The origin of the large differences in $\mathscr{R}^*(47)$ for the two rhodopsin adaptations can be examined in detail by comparing $W = 47$ profiles. As shown in Fig. 1, these deviations are concentrated in a few secondary structures (100-150 (TM2-CP2-TM3) and 290-end (TM7-CP4)). What is most striking is that sum of the river lamprey rhodopsin's $\mathscr{R}^*(47)$ excess fluctuations over the deep sea lamprey rhodopsin's exceeds the reverse sum by a factor of 3.5, a larger factor than the concentration of mutations in TM compared to EC and CP loops. Given the limitations of profiling for fixed $W = 47$, this is persuasive evidence that the main function of the 29 deep sea lamprey rhodopsin mutations is to smooth long wave length hydropathic fluctuations.

**Evolutionary Synchronization**

In elephants and whales a thick (tough white outer eye envelope) sclera is found[11]. Comparison of rhodopsin $\mathscr{R}$*(W) for elephants, whales and humans shows a more complex behavior than river/deep sea lamprey (Table II). The whale/human $\mathscr{R}$*(W) ratio shows whales relatively smoothest at W = 9, where the short-range effects become smaller than long-range ones. This is easily understood: the effects of deep sea pressures for whales are largest at this crossover. Elephants have masses similar to those of smaller whales, but the marine pressures are absent, leaving only residual large mass effects. Moreover, mastodons, Asian and African elephant species diverged about 7 million years ago, roughly the same time primate species (gorilla, human, chimp) diverged[12]. Indeed, $\mathscr{R}$*(W) for the elephant and human rhodopsin agree to within 3% for $1 \leq W \leq 47$ in spite of a factor of 100 difference in body mass[11].

**Non-marine proximate (critical) mammalian evolution**

It is all very well to discuss differences between elephants and whales and humans, but the pharmaceutical industry is much more interested in the small differences between humans, monkeys, cats, mice and rabbits. For non-marine species, human rhodopsin (348 aa) can be used as an absolute benchmark, and the short-range BLAST similarity of these five proximate species' rhodopsin to human rhodopsin is listed in Table III. The evolutionary hierarchy given by BLAST is pretty much as expected: human, monkey, (rabbit, cat), mouse. The correlation coefficients C = |R| of the BLAST scores with $\mathscr{R}^*(W)$ are impressive, as for W = 1, C = 0.54 (compare to conventional "folding"[13], limited at present to less than 50% success for proteins smaller than 150 aa), while for W = 3, already C = 0.86 (excellent!), but the optimal value occurs at W ~ 25 (1 TM length), and here C = 0.96, which is possible only because of SONC.

There is still more than a little mystery to these very high correlation values for very closely related species (BLAST 692-717). Adding dog [BLAST 675] to the list reduces C(25) to 0.92, but adding dog, pig [673] and sheep [662] reduces C(25) to 0.32. Apparently there is something special about rhodopsin in human, monkey, (rabbit, cat), mouse, and perhaps even dog, which is absent in most other mammals. One could even say that these five species define a critically optimized mammalian rhodopsin subfamily.

We can explore these unexpected correlations (comparing BLAST similarities to $\mathscr{R}^*(W)$ resembles comparing apples to oranges) in two ways: (1) use the MZ scale, but compute the roughness as a simple variance with a common protein-wide average, and (2) continue to separate EC, CP and TM regions, but use the short-range KD transference hydropathic scale[4] instead of the long-range MZ scale. The results (Fig. 2) show that with (1) the maximum C = 0.96 seen near W = 25 (TM length) disappears, while C still remains large > 0.93±0.02, while the KD scale shifts the peak in C to ~ W = 10, a medium-range length which may reflect the harmonic average of short-range transference interactions ( W~ 3) and the intrinsic TM length. In practice, since the short-range interactions are already well handled by BLAST-based libraries for specific protein families, the advantages of using $\mathscr{R}^*(W)$ [which here attains the remarkable maximum of $\mathscr{R}^*(25) = 0.96$] to treat long-range interactions separately are obvious.

**Extreme night vision**

Because $\mathscr{R}^*(W)$ is so informative, it is even capable of identifying and explaining secondary aspects of rhodopsin night vision involving mechanical interactions that can supplement and even replace the electronic interactions associated with the ganglion network that transmits electronic signals to the visual cortex. Table IV shows results that suggest that such secondary mechanisms may exist and be supported by mechanical rhodopsin-rod-matrix interactions which have even longer ranges than electronic ones communicated through the ganglion network, which is missing in blind mole rat[14]. The residual eyes of blind mole rat are safely subcutaneous, and although the ganglion network is missing, the mechanical bilateral projection from the retina to the suprachiasmatic nucleus has expanded relative to mouse. This projection could store and transmit mechanical stress signals.

Careful reading of trends in $\mathscr{R}^*(47)/\mathscr{R}^*(25)$ values suggests that the anomalous decrease seen in blind mole rat is present in some other species as well. If we compare bat to cat and mouse, we see that bat $\mathscr{R}^*(47)/$Human - $\mathscr{R}^*(25)/$Human should have been 0.06, but it is actually only 0.02. Genetic analysis has suggested that molecular adaptation of rhodopsin is distinctive in echo locating bats, and it was concluded that the independent evolution of rhodopsin vision in mammals inhabiting low light environments has involved molecular evolution at the sequence level, though most of these changes might not mediate spectral sensitivity directly[15]. As we saw above, typically one has ~ 30 mutations, and only 3 of these determine the spectral sensitivity.

Bovine is by far the most surprising entry in Table IV; similar mammals (sheep, pig, etc.) do not exhibit negative $\mathscr{R}^*(47)$/Human - $\mathscr{R}^*(25)$/Human, and this presents a dilemma. Is it possible that bovine has developed exceptional secondary night vision to supplement nocturnal foraging? Or could bovine (but not sheep, pig) have developed longer range sensitivity[16] for some other reason? The answer to this question lies outside the present analysis. Note that such large smoothing of bovine $\mathscr{R}^*(47)$ is very unlikely to be a statistical accident, as two averages are involved, first over W = 47, and then over the entire 348 aa bovine sequence.

**Conclusions**

While these physiological results are striking, the real value of the results shown in the Tables is much greater. It is clear that the long-range hydropathic hierarchies are functionally much more successful than the standard short-range BLAST multiple alignment hierarchies, and that they can be used to analyze adaptive plasticity and protein network stresses in a wide range of contexts[17,18,19]. The ability to analyze ultra-proximate interspecies differences has been demonstrated for four cases: deep sea marine environment, synchronized evolution, extreme night vision, and five (or six) proximate maximally evolved mammals (including human). The ability to analyze human-mouse-rabbit functional hydro-stress induced differences without adjustable parameters could be most useful in the context of engineering humanized mouse- or rabbit-derived monoclonal Antibodies (mAbs), which are receiving much pharmaceutical attention[20]. For IG structures, one would also use three regions (heavy chain, light chain, and J), and separated hydropathic analysis should be a useful supplement to short length (L = 9) libraries[20].

Readers who are unfamiliar with the general philosophy of power law distributions and critical fluctuations will find many mathematical references at

http://www.cscs.umich.edu/~crshalizi/notebooks/phase-transitions.html

Of course, for proteins we have strong evidence that evolution by plastic adaptation proceeds in a self-similar way for proximate species: a good example is the synchronous elephant-human rhodopsin evolution discussed here. Given the economy and simplicity of hydroanalysis, it appears that many more such examples can be found.

These calculations were expedited greatly by an EXCEL macro devised by N. Voorhoeve.

## References

1. M. A. Moret and G. F. Zebende Phys. Rev. E **75**, 011920 (2007).

2. J. C. Phillips, Phys. Rev. E **80**, 051916 (2009).

3. B. Kobilka and G. F. X. Schertler, Trends Phar. Sci. **2**, 79 (2008).

4. J. C. Phillips, arXiv 1011.4286 (2010).

5. R. J. Trabanino, S. E. Hall, N. Vaidehi, *et al*., Biophys. J. **86**, 1904 (2004).

6. T. Imai, and N. Fujita, Prot. Struc. Func. Bioinfor. **56**, 650 (2004).

7. N. Salamin, R. O. Wuest, S. Lavergne, *et al.*, Trends Ecol. Evolut. **25**, 692 (2010).

8. S. Archer, A. Hope, and J. C. Partridge, Proc. Roy. Soc. (London) B-Biol. Sci. **262**, 289 (1995).

9. H. Zhang, and S. Yokoyama, Gene **191**, 1 (1997).

10. T. Sugawara, H. Imai, M. Nikaido, *et al*., Mol. Biol. Evolu. **27**, 506 (2010).

11. R. F. Burton, J. Zoology **269**, 225 (2006).

12. N. Rohland, A. S. Malaspinas, J. L. Pollack, *et al*., Plos Biol. **5**, 1663 (2007).

13. Y. Zhang, Current Opin. Struc. Biol. **18**, 342 (2008).

14. J. W. H. Janssen, P. H. M. Bovee-Geurts, Z. P. A. Peeters, *et al*., J. Bio. Chem. **275**, 38674 (2000).

15. H. B. Zhao, B. H. Ru, E. C. Teeling, *et al*., Plos One **4**, e8326 (2009).

16. F. Piazza, and Y. H. Sanejouand, Phys. Biol. **6**, 046014 (2009).

17. S. A. Dudley, and J. Schmitt, Amer. Natural. **147**, 445 (1996).

18. B. S. McEwen, Brain Res. SI **886**, 172 (2000).

19. P. D. Gluckman, M. A. Hanson, T. Buklijas, *et al*., Nature Rev. Endocrin. **5**, 401 (2009).

20. M. J. Bernett, S. Karki, G. L. Moore, *et al.*, J. Mol. Biol. **396**, 1474 (2010).

|  | Human | River lamprey | Deep sea lamprey |
|---|---|---|---|
| Uniprot | P08100 | Q90215 | Q90214 |
| BLAST/Human | 1.00 | 0.82 | 0.83 |
| $\mathscr{R}$*(25)/Human | 1.00 | 1.04 | 0.97 |
| $\mathscr{R}$*(47)/Human | 1.00 | 1.59 | 1.10 |

Table I.  Although the deep sea lamprey rhodopsin appears to be little different from the river lamprey rhodopsin as regards short-range BLAST sequence similarity, it is smoother than human rhodopsin at W = 25 = 1 TM length, and almost as smooth as human rhodopsin at W = 47 ~ 2 TM lengths, where the river lamprey is almost as rough as chicken (not shown).

|  | Human (717) | Elephant (677) | Whale(681) |
|---|---|---|---|
| $\mathscr{R}$*(3)/Human | 1.00 | 1.02 | 0.92 |
| $\mathscr{R}$*(9)/Human | 1.00 | 1.00 | 0.72 |
| $\mathscr{R}$*(25)/Human | 1.00 | 1.03 | 0.82 |
| $\mathscr{R}$*(47)/Human | 1.00 | 1.03 | 1.31 |

Table II.  Both elephant and whale rhodopsin roughness reach their smallest values (relative to human rhodopsin roughness) at W = 9, but whale roughness varies much more widely with averaging length.  The whale roughness is supposed to reflect high deep sea pressures, while the elephant roughness may reflect residual internal pressures associated with its large body and eye mass.  BLAST against human rhodopsin in parentheses.

BLAST   $\mathscr{R}$*(3)/Human   $\mathscr{R}$*(9)/Human   $\mathscr{R}$*(25)/Human   $\mathscr{R}$*(47)/Human

| | | | | | |
|---|---|---|---|---|---|
| Correl. (|R|) | | 0.86 | 0.92 | 0.96 | 0.92 |
| Human | 717 | 1.00 | 1.00 | 1.00 | 1.00 |
| Monkey | 705 | 1.01 | 1.02 | 1.07 | 1.05 |
| Rabbit | 701 | 1.03 | 1.04 | 1.07 | 1.09 |
| Cat | 701 | 1.04 | 1.06 | 1.11 | 1.16 |
| Mouse | 692 | 1.04 | 1.11 | 1.14 | 1.23 |

Table III.  BLAST similarity scores (compared to human), relative roughness scores $\mathscr{R}$*(W) for W = 3, 9, 25 and 47 for five mammalian species.  Human roughness is smallest, and the interspecies differences increase with increasing W, with the best correlation to BLAST occurring at the transmembrane length W = 25.  Note that rabbit (prey) is smoother than cat (predator) for large W, although the BLAST scores are equal.

| | BLAST | $\mathscr{R}$*(25)/Human | $\mathscr{R}$*(47)/Human | Uniprot |
|---|---|---|---|---|
| Human | 717 | 1.00 | 1.00 | P08100 |
| Bat | 695 | 1.13 | 1.15 | D2WK06 |
| Blind mole rat | 686 | 1.04 | 0.98 | Q9ERF2 |
| Bovine | 681 | 1.10 | 0.98 | P02699 |

Table IV.   Species with anomalously small $\mathscr{R}$*(47)/Human and $\mathscr{R}$*(47)/$\mathscr{R}$*(25) values. These may be connected to a secondary visual mechanism (see text).  This anomaly does not occur in blind cave fish (P41590).

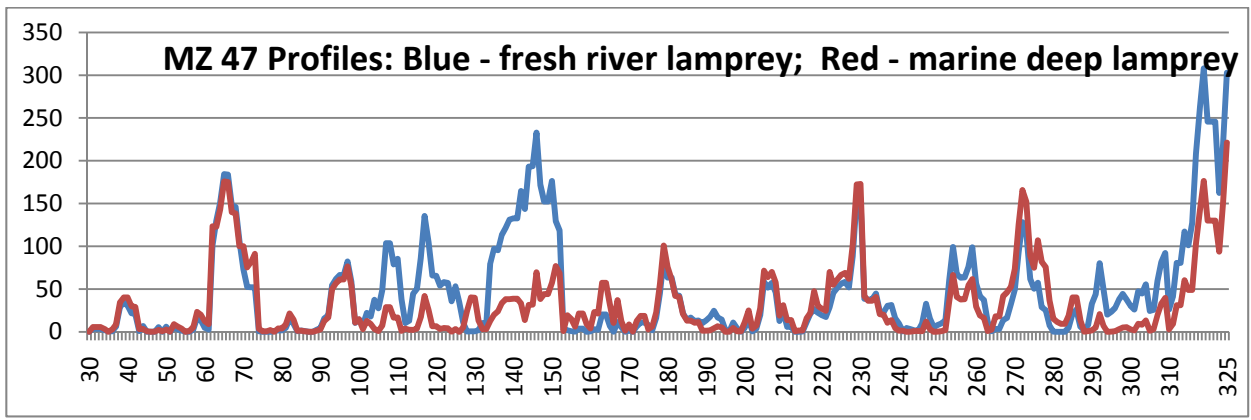

Fig. 1. MZ hydropathic profiles averaged over a W = 47 window. Note the dramatic smoothing of the marine deep lamprey, especially in the 100-150 (TM2-CP2-TM3) and 290-end (TM7-CP4) regions. Amino acid numbering as in Uniprot.

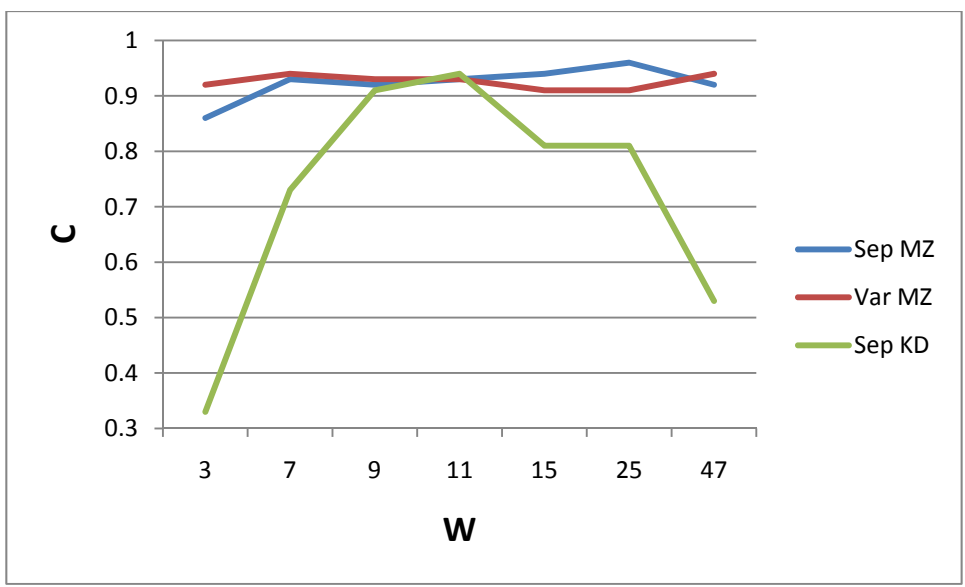

Fig. 2. Rhodopsin correlation C of roughening $\mathscr{P}*$(W) with BLAST similarity to human of five species (humans, monkeys, cats, mice and rabbits).